\documentclass[aps,print]{revtex4}
\usepackage{amsfonts}
\usepackage{amsmath}
\usepackage{amssymb}
\usepackage{graphicx}
\usepackage{color}

\begin{document}

\title{Ring-shaped quantum droplets with hidden vorticity in a
radially-periodic potential}
\author{Bin Liu$^{1,2}$, Xiaoyan Cai$^{1}$, Xizhou Qin$^{1,2}$, Xunda Jiang$%
^{1,2}$}
\author{Jianing Xie$^{1,2}$}
\email{xiejianingfs@126.com}
\author{Boris A. Malomed$^{3,4}$}
\author{Yongyao Li$^{1,2}$}
\affiliation{$^{1}$School of Physics and Optoelectronic Engineering, Foshan University,
Foshan 528000, China}
\affiliation{$^{2}$Guangdong-Hong Kong-Macao Joint Laboratory for Intelligent Micro-Nano
Optoelectronic Technology, Foshan University, Foshan 528000, China}
\affiliation{$^{3}$Department of Physical Electronics, School of Electrical Engineering,
Faculty of Engineering, and Center for Light-Matter Interaction, Tel Aviv
University, Tel Aviv 69978, Israel}
\affiliation{$^{4}$Instituto de Alta Investigaci\'{o}n, Universidad de Tarapac\'{a},
Casilla 7D, Arica, Chile}

\begin{abstract}
We study the stability and characteristics of two-dimensional (2D) circular
quantum droplets (QDs) with embedded hidden vorticity (HV), i.e., opposite
angular momenta in two components, formed by binary Bose-Einstein
condensates (BECs) trapped in a radially-periodic potential. The system is
modeled by the Gross-Pitaevskii (GP) equations with the Lee-Huang-Yang (LHY)
terms, which represent the higher-order self-repulsion induced by quantum
fluctuations around the mean-field state, and a potential which is a
periodic function of the radial coordinate. Ring-shaped QDs with high
winding numbers (WNs) of the HV type, which are trapped in particular
circular troughs of the radial potential, are produced by means of the
imaginary-time-integration method. Effects of the depth and period of the
potential on these QD states are studied. The trapping capacity of
individual circular troughs is identified. Stable compound states in the
form of nested multiring patterns are constructed too, including ones with
WNs of opposite signs. The stably coexisting ring-shaped QDs with different
WNs can be used for the design of BEC-based data-storage schemes.
\end{abstract}

\maketitle

\section{Introduction}

Bose-Einstein condensates (BECs) have long been an ideal platform for
studying various physical problems, such as stability and dynamics of
self-localized states \cite{BEC_WB,PRA97_063607,cbp_zlc,cbp2,cpl_zlc,cpl1}.
It is well known that the cubic self-attraction gives rise to stable bright
solitons only in effectively one-dimensional (1D) systems. Actually, the
natural sign of the mean-field (MF) nonlinearity for BEC is self-repulsive,
while it may be switched into attraction, which is necessary for
self-trapping of solitons, by means of the Feshbach resonance \cite{Feshbach}%
. However, in 2D and 3D settings, matter-wave solitons are destabilized by
the occurrence of the, respectively, critical and supercritical collapse in
the same dimension \cite{Towns_soliton,Malomed2005,Berge1998}. Therefore,
stabilization of multidimensional solitons, especially vorticity-carrying
ones, is a well-known challenging problem \cite{book}. The most common
method for the solution of the problem is to modify the nonlinearity. This
is possible with the help of quadratic self-interaction \cite{Xliu2000},
competitive self-focusing and defocusing terms \cite%
{Mihalache2001,WYYDN90,ND93_2379,ND91_757}, saturable self-attraction \cite%
{Segev1994}, and nonlocal nonlinearity \cite%
{Peccianti2002,Pedri2005,Tikonenkow2008,Jiasheng,Maucher2011}. Other methods
use spin-orbit coupling in binary BECs \cite{Stanescu2008,Han Pu,Ben
Li,lbNJP}, the harmonic-oscillator trapping potential \cite%
{Harmonic1,LZH_PRE} or spatially periodic (lattice) potentials \cite{SR}.

Recently, a new kind of quantum matter, in the form of quantum droplets
(QDs), has been experimentally created in dipolar \cite%
{Schmitt2016,Chomaz2016} and binary bosonic gases \cite{Cabrera2018}. QDs
rely on their intrinsic nonlinearity to form stable self-localized states.
Originally, the formation of QDs was predicted on the basis of the balance
between the MF self-attraction and the repulsion induced by quantum
fluctuations beyond the MF approximation \cite{Petrov2015,Petrov2016}. The
latter effect is represented by the Lee-Huang-Yang (LHY) \cite{LHY}
corrections added to the corresponding Gross-Pitaevskii (GP) equations. The
predictions demonstrate that the LHY terms take different forms in 1D, 2D
and 3D systems, being self-repulsive in the 2D and 3D cases. Studies of QDs have drawn much interest in very
different contexts \cite%
{PRA101_051601,PRA98_033612,PRA102_043302,PRA102_023318,PRL126_025301,PRResearch2_033522,PRE102_062217,FOP16_32201_LYY,PRA99_053602,CXL_PRA2018,PRL120_160402,PRResearch2_043074,PRL126_244101,PRResearch3_033247,CSF_WHC,FOP_GMY,FOP_Boris,CSF2022_ZJF,Nonlinear Dyn. Zhou 2022,Front. Phys. Hu 2022,CSF Zhou 2021,NJP1,NJP3,NJP5,NJP6,NJP7,NJP8,NJP9,cpb5,cpb_yw,cpb_lxj,PRl_122_090401,FOP_LGL,photonicis_YAW}%
. In particular, stable QDs with embedded vorticity have been predicted in
2D and 3D forms \cite{PRA_LYY_2DQD2018,KTS2018}, including semidiscrete
vortex QDs in arrays of coupled 1D matter-wave conduits \cite{PRL123_133901}%
. However, stability conditions for vortex QDs become increasingly stricter
with the increase of their winding number (WN, alias the topological
charge). Therefore, searching for experimentally relevant settings that
promote the existence of such stable topological modes is a relevant
problem. QDs in an effectively 1D system have been considered too, where the
quadratic LHY term added to the GP equation is self-attractive, on the
contrary to its sign in the 3D and 2D settings \cite{Zouzheng,CSF_zfy,ND_DLW}%
.

Another possibility for the stabilization of nonlinear vortex modes has been
revealed by studies of BEC systems under the action of spatially periodic
and quasiperiodic lattice potentials \cite%
{RFZ2,ZXFPRA95,WJG,BEC_OL,cpl_dcq,cpl_xsl,cpb_wlh,cpb1,cpb3,FOP_ZYY}. It has
been predicted that periodic potentials not only help to stabilize
self-bound vortex modes, but also alter their formation dynamics. The same
pertains to localized states with \textit{hidden vorticity} (HV), which are
defined as two-component states with equal norms and opposite angular
momenta (i.e., opposite vorticities) of their components \cite{Brtka}. HV
states in BECs in a rotating double-well potential were considered in Ref.
\cite{HV_WB,HV_pla}.

As an experimentally\ relevant means of supporting stable self-trapped
vortex modes, especially ones with high topological charges, it is more
natural to use radial potential, which, unlike the spatially periodic ones,
conserve the angular momentum. In particular, Ref. \cite{PRA_HCQ2017}
reported stable vortex solitons with WN $S=11$ maintained by the {radial
potential in a dipolar BEC with repulsive long-range dipole-dipole
interactions, and Ref. \cite{NJP_LB2022} reported stable narrow vortex-ring
QDs with high values of $S$ trapped in particular circular troughs of the
radial potential. Further, Ref. \cite{NJP_LB2022} reported stable compound
states in the form of multiple mutually nested concentric rings. However,
such complexes are stable only when the inner and outer annular components
of the nested complexes are widely separated, hence they practically do not
interact.

The objective of this work is to demonstrate that stable 2D ring-shaped QDs
of the HV type can be created and controlled in binary BECs with contact
interactions, using radially-periodic potentials. The dynamics of the system
are modeled by coupled GP equations including the LHY terms and
radial-periodic potential. The corresponding annular (ring-shaped) HV QDs
are trapped in particular circular troughs of the radial potential. The
effects of the modulation depth and period of the potential states are
systematically studied. Throughout this investigation, stability areas are
identified versus the number of the trapping radial trough and WN of the HV
states. Then, nested complexes of annular QDs that carry different WNs are
addressed. In particular, the results offer a possibility to design a new
encoding device that makes use of coexisting HV ring-shaped QDs with
different WNs numbers for storing data components.

The rest of the paper is structured as follows. The model is introduced in
Sec. II, numerical findings and some analytical estimates regarding 2D
ring-shaped HV QDs are summarized in Sec. III, and multi-ring nested
patterns with different WNs in concentric circular troughs are the focus of
Sec. IV. The work is concluded by Sec. V.

\section{The model}

Following Refs. \cite{PRL120_235301, CNSNS_LZD}, we assume that QDs, which
are formed by binary BEC, are relatively loosely confined in the transverse
direction $z$, with confinement size $a_{\bot }$ $\sim $ a few microns,
relative to the system's plane $\left( x,y\right) $. The corresponding
system of coupled GP equations for wave functions $\Psi _{1,2}$ of two
components of the binary condensate, including the LHY correction (quartic
self-repulsion) can be written as %
\begin{eqnarray}
i\hbar \frac{\partial }{\partial t}\Psi _{j} &=&\left( -\frac{\hbar ^{2}}{2m}%
\nabla _{\text{3D}}^{2}+V(\mathbf{r})+\frac{\delta E_{\text{total}}}{\delta
n_{j}}\right) \Psi _{j},  \label{GPE} \\
E_{\text{total}} &=&E_{\text{MF}}+E_{\text{LHY}}  \label{E-total}
\end{eqnarray}%
where $j=1,2$, $\nabla _{\text{3D}}^{2}=\partial _{z}^{2}+\nabla _{\text{2D}%
}^{2}$, and $\nabla _{\text{2D}}^{2}\equiv \partial _{r}^{2}+r^{-1}\partial
_{r}+r^{-2}\partial _{\theta }^{2}$ is the 2D Laplacian written in the polar
coordinates, while $V(\mathbf{r})$ is the potential. The total energy
includes the mean-field and LHY terms,

\begin{eqnarray}
E_{\text{MF}} &=&\int_{-\infty }^{+\infty }dz\int_{0}^{\infty
}rdr\int_{0}^{2\pi }d\theta \left( \frac{1}{2}%
g_{11}n_{1}^{2}+g_{12}n_{1}n_{2}+\frac{1}{2}g_{22}n_{2}^{2}\right) ,
\label{E-MF} \\
E_{\text{LHY}} &=&\frac{8m^{3/2}}{15\pi ^{2}\hbar ^{3}}\int_{-\infty
}^{+\infty }dz\int_{0}^{\infty }rdr\int_{0}^{2\pi }d\theta \left(
g_{11}n_{1}+g_{22}n_{2}\right) ^{5/2},  \label{E-LHY}
\end{eqnarray}%
where the $n_{j}=|\Psi _{j}|^{2}$ are the densities, $g_{11}$, $g_{22}$ and $%
g_{12}$ are strengths of the MF self- and cross-interactions of the two
species, which satisfy constraints $g_{11}$, $g_{22}>0$ and $g_{12}+\sqrt{%
g_{11}g_{22}}<0$ \cite{Petrov2015}. Here, we consider the case
of equal masses, i.e., $m_{1}=m_{2}=m$, and $g_{i,j}=4\pi a_{i,j}/m$, where $%
a_{11,12}$ and $a_{12}$ are the intraspecies and interspecies scattering
length, respectively. It is well known that feshbach resonances are an
important tool for extracting information about interactions between atoms.
In addition, they provide a way to change the scattering length almost at
will \cite{book_BEC}.{\emph{\ }}If we consider, as usual, the symmetric
system, with $g=g_{11}=g_{22}>0$ and $g_{12}=-\left( g+\delta g\right) $,
with $\delta g>0$, the system of the GP equations is obtained in the form of

\begin{align}
i\hbar \frac{\partial \Psi _{1}}{\partial t}& =-\frac{\hbar ^{2}}{2m}\nabla
_{\text{3D}}^{2}\Psi _{1}+g\left( \left\vert \Psi _{1}\right\vert
^{2}-\left\vert \Psi _{2}\right\vert ^{2}\right) \Psi _{1}-\delta
g\left\vert \Psi _{2}\right\vert ^{2}\Psi _{1}+\frac{4g^{5/2}m^{3/2}}{3\pi
^{2}\hbar ^{3}}\left( \left\vert \Psi _{1}\right\vert ^{2}+\left\vert \Psi
_{2}\right\vert ^{2}\right) ^{3/2}\Psi _{1}+V\left( r\right) \Psi _{1},
\notag \\
i\hbar \frac{\partial \Psi _{2}}{\partial t}& =-\frac{\hbar ^{2}}{2m}\nabla
_{\text{3D}}^{2}\Psi _{2}+g\left( \left\vert \Psi _{2}\right\vert
^{2}-\left\vert \Psi _{1}\right\vert ^{2}\right) \Psi _{2}-\delta
g\left\vert \Psi _{1}\right\vert ^{2}\Psi _{2}+\frac{4g^{5/2}m^{3/2}}{3\pi
^{2}\hbar ^{3}}\left( \left\vert \Psi _{1}\right\vert ^{2}+\left\vert \Psi
_{2}\right\vert ^{2}\right) ^{3/2}\Psi _{2}+V\left( r\right) \Psi _{2}.
\label{GPE_3D}
\end{align}%
Adopting, as said above, the potential $V\left( \mathbf{r}\right) $ as

\begin{equation}
V\left( \mathbf{r}\right) =V\left( x,y\right) +\frac{1}{2}m\omega
_{z}^{2}z^{2},  \label{V_3D}
\end{equation}%
one can carry out the 3D $\rightarrow $ 2D reduction by substituting
\begin{equation}
\Psi _{1,2}\left( x,y,z,t\right) =\Phi _{1,2}\left( x,y,t\right) \left(
\frac{m\omega _{z}}{\pi \hbar }\right) ^{1/4}\exp \left( -\frac{m\omega _{z}%
}{2\hbar }z^{2}-i\frac{\omega _{z}}{2}t\right) ,  \label{tran_3D-2D}
\end{equation}%
into Eqs. (\ref{GPE_3D}) and averaging the equations in the $z$-direction,
which yields
\begin{align}
i\hbar \frac{\partial \Phi _{1}}{\partial t}& =-\frac{\hbar ^{2}}{2m}\nabla
_{\text{2D}}^{2}\Phi _{1}+\left( \frac{m\omega _{z}}{3\pi \hbar }\right)
^{1/2}\left[ g\left( \left\vert \Phi _{1}\right\vert ^{2}-\left\vert \Phi
_{2}\right\vert ^{2}\right) -\delta g\left\vert \Phi _{2}\right\vert ^{2}%
\right] \Phi _{1}  \notag \\
& +\left( \frac{m\omega _{z}}{\pi \hbar }\right) ^{3/4}\frac{2g^{5/2}m^{3/2}%
}{3\pi ^{2}\hbar ^{3}}\left( \left\vert \Phi _{1}\right\vert ^{2}+\left\vert
\Phi _{2}\right\vert ^{2}\right) ^{3/2}\Phi _{1}+V\left( x,y\right) \Phi
_{1},  \notag \\
i\hbar \frac{\partial \Phi _{2}}{\partial t}& =-\frac{\hbar ^{2}}{2m}\nabla
_{\text{2D}}^{2}\Phi _{2}+\left( \frac{m\omega _{z}}{3\pi \hbar }\right)
^{1/2}\left[ g\left( \left\vert \Phi _{2}\right\vert ^{2}-\left\vert \Phi
_{1}\right\vert ^{2}\right) -\delta g\left\vert \Phi _{1}\right\vert ^{2}%
\right] \Phi _{2}  \notag \\
& +\left( \frac{m\omega _{z}}{\pi \hbar }\right) ^{3/4}\frac{2g^{5/2}m^{3/2}%
}{3\pi ^{2}\hbar ^{3}}\left( \left\vert \Phi _{1}\right\vert ^{2}+\left\vert
\Phi _{2}\right\vert ^{2}\right) ^{3/2}\Phi _{2}+V\left( x,y\right) \Phi
_{2}.  \label{GP_2D_unnormalized}
\end{align}%
Furthermore, the radially-periodic potential is taken as \cite{BBB}%
\begin{eqnarray}
V\left( x,y\right)  &=&\frac{\hbar ^{2}k_{r}^{2}}{2m}\cos ^{2}\left( \frac{%
\pi }{D_{0}}r\right) ,  \label{V_2D} \\
k_{r} &=&\frac{2\pi }{\lambda },r=\sqrt{x^{2}+y^{2}},  \notag
\end{eqnarray}%
where $D_{0}$ is the radial period of the potential, and $\lambda $ is the
laser wavelength of potential-building. Such a potential can be
readily imposed by a broad laser beam passed through a properly designed
phase plate \cite{phase-plate}. By means of rescaling with length unit $r_{0}
$ and following transformations}%
\begin{eqnarray}
t^{\prime } &=&\frac{\hbar }{mr_{0}^{2}}t,x^{\prime }=x/r_{0},y^{\prime
}=y/r_{0},D=D_{0}/r_{0},\psi _{1,2}=r_{0}\Phi _{1,2},V\left( x^{\prime
},y^{\prime }\right) =\frac{mr_{0}^{2}}{\hbar ^{2}}V\left( x,y\right) ,
\notag \\
g^{^{\prime }} &=&\frac{m}{\hbar ^{2}}\left( \frac{m\omega _{z}}{3\pi \hbar }%
\right) ^{1/2}g,\delta g^{^{\prime }}=\frac{m}{\hbar ^{2}}\left( \frac{%
m\omega _{z}}{3\pi \hbar }\right) ^{1/2}\delta g,\Gamma =\frac{m}{r_{0}\hbar
^{2}}\frac{2g^{5/2}m^{3/2}}{3\pi ^{2}\hbar ^{3}}\left( \frac{m\omega _{z}}{%
\pi \hbar }\right) ^{3/4}  \label{Nondimensional}
\end{eqnarray}%

Eq. (\ref{GP_2D_unnormalized}) is cast in the normalized form,

\begin{align}
i\frac{\partial \psi _{1}}{\partial t}& =-\frac{1}{2}\nabla _{\text{2D}%
}^{2}\psi _{1}+g\left( \left\vert \psi _{1}\right\vert ^{2}-\left\vert \psi
_{2}\right\vert ^{2}\right) \psi _{1}-\delta g\left\vert \psi
_{2}\right\vert ^{2}\psi _{1}+\Gamma \left( \left\vert \psi _{1}\right\vert
^{2}+\left\vert \psi _{2}\right\vert ^{2}\right) ^{3/2}\psi _{1}+V_{0}\cos
^{2}\left( \frac{\pi }{D}r\right) \psi _{1},  \notag \\
i\frac{\partial \psi _{2}}{\partial t}& =-\frac{1}{2}\nabla _{\text{2D}%
}^{2}\psi _{2}+g\left( \left\vert \psi _{2}\right\vert ^{2}-\left\vert \psi
_{1}\right\vert ^{2}\right) \psi _{2}-\delta g\left\vert \psi
_{1}\right\vert ^{2}\psi _{2}+\Gamma \left( \left\vert \psi _{1}\right\vert
^{2}+\left\vert \psi _{2}\right\vert ^{2}\right) ^{3/2}\psi _{2}+V_{0}\cos
^{2}\left( \frac{\pi }{D}r\right) \psi _{2},  \label{GPE_2D}
\end{align}%
where $V_{0}$ and $D$ are the accordingly rescaled depth and radial period
of the potential. Localized solutions of Eq. (\ref{GPE_2D}) are
characterized by their 2D norm,%
\begin{equation}
N_{\text{2D}}=\int_{0}^{\infty }rdr\int_{0}^{2\pi }d\theta \left( \left\vert
\psi _{1}\right\vert ^{2}+\left\vert \psi _{2}\right\vert ^{2}\right) .
\label{norm}
\end{equation}

According to the ref. \cite{book_BEC,phase-plate,BBB}, the
nonlinear coefficient and LHY corrections can be adjusted by Feshbach
resonances, and the wavelength and incident angle of the laser are also
adjusted. In this work, we refer to parameters for BEC of $^{39}$K atoms,
the potential-building laser wavelength $\lambda =1064$ nm,
transverse-confining frequency $\omega _{z}=2\pi \times 200$ Hz, and
rescaling length $r_{0}=0.08\lambda $, which yields $g=0.057$, $\delta g=g/10
$, and $\Gamma =0.0049$. A typical simulation time, for which the results
are reported below, $t=20000$, is equivalent to $\approx 100$ ms in physical
units, which is a sufficiently large time for the experiment.%
While, according to Eq. (\ref{tran_3D-2D}) and (\ref{norm}), the total atoms
is $N_{\text{2D}}$.

The objective of the present work is to predict stable ring-shaped QDs of
the HV type as solutions of Eq. (\ref{GPE_2D}). They are defined by fixing
WNs and chemical potentials of the two components as $S_{2}=-S_{1}=-S$ and $%
\mu _{2}=\mu _{1}=\mu $ \cite{Brtka}. Thus, the seek for stationary HV
solutions in the form of
\begin{align}
\psi _{1}\left( r,\theta ,t\right) & =\phi _{1}\left( r\right) e^{-i\mu
t+iS\theta },  \notag \\
\psi _{2}\left( r,\theta ,t\right) & =\phi _{2}\left( r\right) e^{-i\mu
t-iS\theta },  \label{Solution}
\end{align}%
where real functions $\phi _{1,2}$ obey radial equations:
\begin{align}
\mu \phi _{1}& =-\frac{1}{2}\left( \frac{d^{2}}{dr^{2}}+\frac{1}{r}\frac{d}{%
dr}-\frac{S^{2}}{r^{2}}\right) \phi _{1}+g\left( \left\vert \phi
_{1}\right\vert ^{2}-\left\vert \phi _{2}\right\vert ^{2}\right) \phi
_{1}-\delta g\left\vert \phi _{2}\right\vert ^{2}\phi _{1}+\Gamma \left(
\left\vert \phi _{1}\right\vert ^{2}+\left\vert \phi _{2}\right\vert
^{2}\right) ^{3/2}\phi _{1}+V_{0}\cos ^{2}\left( \frac{\pi }{D}r\right) \phi
_{1},  \notag \\
\mu \phi _{2}& =-\frac{1}{2}\left( \frac{d^{2}}{dr^{2}}+\frac{1}{r}\frac{d}{%
dr}-\frac{S^{2}}{r^{2}}\right) \phi _{2}+g\left( \left\vert \phi
_{2}\right\vert ^{2}-\left\vert \phi _{1}\right\vert ^{2}\right) \phi
_{2}-\delta g\left\vert \phi _{1}\right\vert ^{2}\phi _{2}+\Gamma \left(
\left\vert \phi _{2}\right\vert ^{2}+\left\vert \phi _{1}\right\vert
^{2}\right) ^{3/2}\phi _{2}+V_{0}\cos ^{2}\left( \frac{\pi }{D}r\right) \phi
_{2}.  \label{radial_equation}
\end{align}

Stationary ring-shaped QD vortices trapped in radial potential troughs with
numbers O$_{\text{n}}=1,2,3...$ can be produced by means of the
imaginary-time-propagation (ITP) method \cite{ITP1,ITP2}. It was applied to
Eq. (\ref{GPE_2D}) with the Gaussian initial guess,
\begin{eqnarray}
\psi _{10}\left( r,\theta \right) &=&C_{1}\exp \left[ -\alpha _{1}\left(
r-r_{\text{n}}\right) ^{2}+iS\theta \right] ,  \notag \\
\psi _{20}\left( r,\theta \right) &=&C_{2}\exp \left[ -\alpha _{2}\left(
r-r_{\text{n}}\right) ^{2}-iS\theta \right] ,  \label{initial_guess}
\end{eqnarray}%
where $\alpha _{1,2}>0$ and $C_{1,2}$ are real constants, and
\begin{equation}
r_{\text{n}}=\left( \text{O}_{\text{n}}-1/2\right) D,  \label{Rn}
\end{equation}%
is the radial coordinate of the bottom point of the given trough. Then, the
stability of the stationary ring-shaped QDs was analyzed by means of the
linearized Bogoliubov -- de Gennes (BdG) equations for perturbed wave
functions,%
\begin{align}
\psi _{_{1}}(x,y,t)& =\left[ \phi _{1}(x,y)+\varepsilon w_{1}(x,y)e^{\lambda
t+im\theta }+\varepsilon v_{1}^{\ast }(x,y)e^{\lambda ^{\ast }t-im\theta }%
\right] e^{-i\mu t+iS\theta },  \notag \\
\psi _{_{2}}(x,y,t)& =\left[ \phi _{2}(x,y)+\varepsilon w_{2}(x,y)e^{\lambda
t-im\theta }+\varepsilon v_{2}^{\ast }(x,y)e^{\lambda ^{\ast }t+im\theta }%
\right] e^{-i\mu t-iS\theta },  \label{pert}
\end{align}%
where $w_{1,2}(x,y)$, $v_{1,2}(x,y)$ and $\lambda $ are eigenmodes and
instability growth rate corresponds to an integer azimuthal index $m$ of the
perturbation with infinitesimal amplitude $\varepsilon $, and $\ast $ stands
for the complex conjugate. The substitution of ansatz (\ref{pert}) in Eq. (%
\ref{GPE_2D}) and linearization leads to the following system of the BdG
equations:%
\begin{equation}
\left(
\begin{array}{cccc}
A_{11} & A_{12} & A_{13} & A_{14} \\
A_{21} & A_{22} & A_{23} & A_{24} \\
A_{31} & A_{32} & A_{33} & A_{34} \\
A_{41} & A_{42} & A_{43} & A_{44}%
\end{array}%
\right) \left(
\begin{array}{c}
w_{1} \\
v_{1} \\
w_{2} \\
v_{2}%
\end{array}%
\right) =i\lambda \left(
\begin{array}{c}
w_{1} \\
v_{1} \\
w_{2} \\
v_{2}%
\end{array}%
\right)  \label{eig_eq}
\end{equation}%
with matrix elements $A_{11}\sim $ $A_{44}$ given in the Appendix.

Numerically solving the linearized equations produces a spectrum of
eigenfrequencies $\lambda $, the stability condition being that the entire
spectrum of $\lambda $ must be pure imaginary \cite{Agrawal,VK}. The so
predicted stability of the stationary states was then verified by direct
simulations of the perturbed evolution, using the fast-Fourier-transform
method.\newline

\section{Single-ring states}

\begin{figure}[tbp]
{\includegraphics[width=0.8\columnwidth]{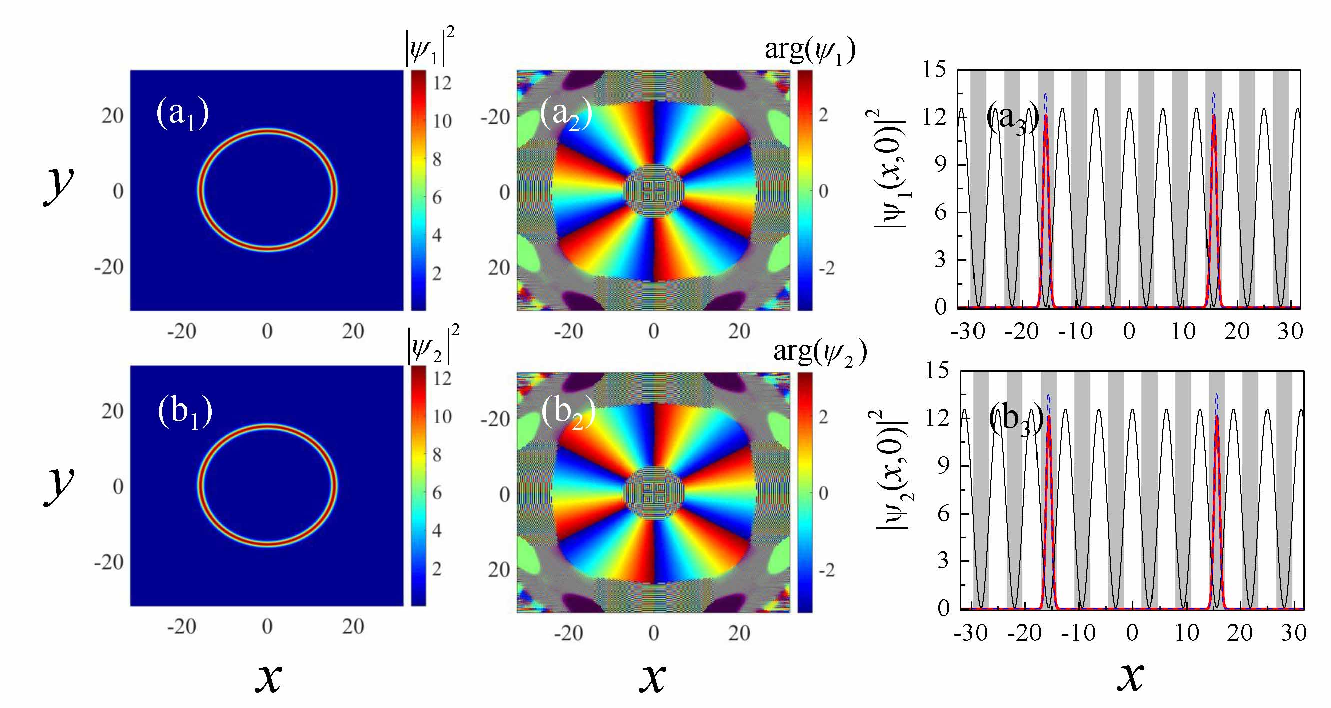}}
\caption{Typical examples of stable ring-shaped QDs of the HV type trapped
in the trough with O$_{\text{n}}=3$. (a$_{1}$,b$_{1}$) Density patterns of
the two components with $N=3000$, $V_{0}=12.63$ and $S_{1,2}=\pm 6$. (a$_{2}$%
,b$_{2}$) The corresponding phase patterns. The red curves in (a$%
_{3}$,b$_{3} $) correspond to the cross section, $|\protect\psi %
_{1,2}(x,0)|^{2}$, along $y=0$, in (a$_{1}$,b$_{1}$). Here, the blue dashed
lines represent, for the comparison's sake, the result for $V_{0}=18.95$,
other parameters remain the same. The black lines represent the
axisymmetric radially-periodic potential with depth $V_{0}=12.63$\ and
period $D=6.25$.}
\label{example}
\end{figure}

Stationary ring-shaped QDs of the HV type, trapped in radial potential
troughs with O$_{\text{n}}=1,2,3$, and $4$, were produced by means of the
ITP method. Typical examples of a numerically constructed QDs with $%
S_{1,2}=\pm 6$, which are trapped in the trough with O$_{\text{n}}=3$, are
displayed in Fig. \ref{example}(a$_{1}$) and (b$_{1}$). This situation
corresponds to the largest WN $\left\vert S_{1}\right\vert =\left\vert
S_{2}\right\vert $ that admits stability for O$_{\text{n}}=3$, as shown
below in Fig. \ref{MU+D}(c). HV phase patterns are shown in Figs. \ref%
{example}(a$_{2}$) and (b$_{2}$), which confirm the condition $S_{2}=-S_{1}$%
. The density cross sections, $|\psi _{1,2}(x,0)|^{2}$, which are displayed
in Fig. \ref{example}(a$_{3}$) and (b$_{3}$), corroborate the placement of
the QDs in the third potential trough. Here, the norm is $N=3000$, and the
radial period and depth of the potential in Eq. (\ref{GPE_2D}) are $D=6.25$
and $V_{0}=12.63$, respectively. While the inner radius of HV QDs in the 2D
free space increases with the increase of the WN and total norm, the radius
of the HV QDs trapped the radially-periodic potential stays nearly constant,
being determined by $r_{\text{n}}$, see Eq. (\ref{Rn}).
\begin{figure}[tbp]
{\includegraphics[width=0.6\columnwidth]{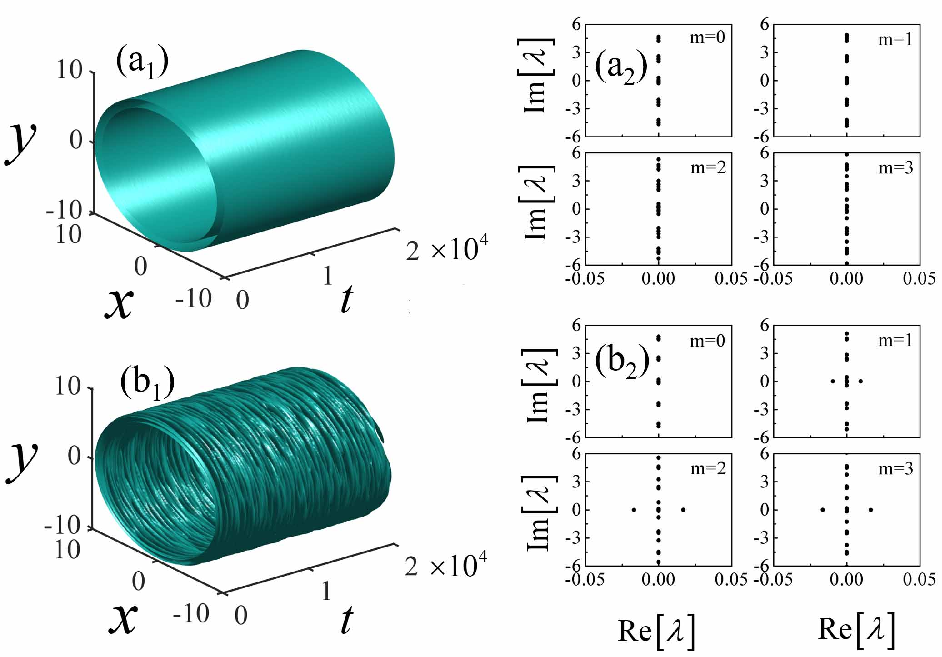}}
\caption{(a$_{1}$) Direct simulations of the $\protect\psi _{1}$ component
of a stable HV QD with $S_{1,2}=\pm 2$ and $N=1500$. (a$_{2}$) Perturbation
eigenvalues for the same mode, for azimuthal indices $m=0,1,2,$and $3$. (b$%
_{1}$) The same as in (a$_{1}$), but for an unstable mode with $N=150$. (b$%
_{2}$) Perturbation eigenvalues for this mode, with azimuthal indices $%
m=0,1,2,$and $3$. The ring-shaped modes displayed in this figure are trapped
in the potential trough with O$_{\text{n}}=2$, other parameters being the
same as in Fig. \protect\ref{example}. {\color{red}Here, $1\%$ random noise
is added with inital ansatz $\protect\psi _{_{1,2}}(x,y,t=0)=\left[ 1+%
\protect\varepsilon ^{^{\prime }}f(x,y)\right] \cdot \protect\phi %
_{1,2}(x,y) $. Where $\protect\varepsilon ^{^{\prime }}=1\%$ and $f(x,y)$ is
a random function with the value range $[0,1]$, that can be generated by
the \textquotedblleft rand\textquotedblright\ function. }}
\label{trans_unstable_stable}
\end{figure}

The stability of the ring-shaped HV QDs was verified by direct simulations
of the perturbed evolution and also through eigenvalues produced by the BdG
equations (\ref{eig_eq}) for small perturbations. Fig. \ref%
{trans_unstable_stable} shows typical examples of stable and unstable QDs
with WNs $S_{1,2}=\pm 2$ trapped in the potential trough with O$_{\text{n}%
}=2 $, for different values of the total norm, $N$. Here, we present direct
simulations and the perturbation eigenvalues with different azimuthal
indices $m$, see Eq. (\ref{pert}). Fig. \ref{trans_unstable_stable}(a$_{1}$)
shows direct simulations of the $\psi _{_{1}}$ component of a stable
ring-shaped HV QD with $1\%$ random noise, and $N=1500$. Further, Figs. \ref%
{trans_unstable_stable}(b$_{1}$) show the same for an unstable HV\ QD, here $%
N=150$. The latter result may be interpreted as fragmentation of the
original azimuthally uniform state as a result of its modulational
instability. Further, Figs. \ref{trans_unstable_stable}(a$_{2}$) and (b$_{2}$%
) show the perturbation eigenvalues for the corresponding modes with $N=1500$
and $N=150$, respectively, for azimuthal indices $m=0$, $1$, $2$, and $3$.
The results demonstrate that the results of direct simulations are
consistent with the prediction of the linear-stability analysis.

\begin{figure}[tbp]
{\includegraphics[width=0.8\columnwidth]{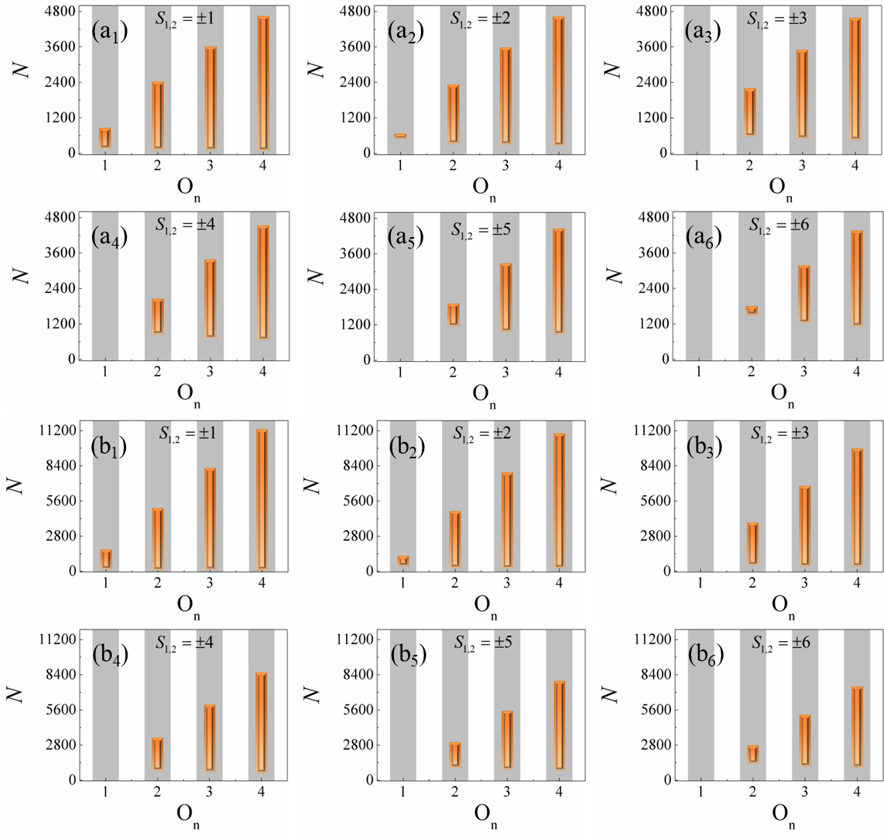}}
\caption{In panels (a$_{1}$)-(a$_{6}$) and (b$_{1}$)-(b$_{6}$), orange bars
represent stability intervals for HV ring-shaped QDs with $S_{1,2}=\pm 1,\pm
2,\pm 3,\pm 4,\pm 5$ and $\pm 6$, respectively, trapped in radial-potential
troughs with numbers O$_{\text{n}}=1,2,3$ and $4$. The depth of the
potential is $V_{0}=12.63$ in panels (a$_{1}$)-(a$_{6}$) and $V_{0}=18.95$
in (b$_{1}$)-(b$_{6}$). The radial period is $D=6.25$.}
\label{stability_N}
\end{figure}

The results of the numerical analysis of the stability of the HV ring-shaped
QDs with different WNs are summarized in Fig. \ref{stability_N} for
different values of the potential depth, \textit{viz}., $V_{0}=12.63$ and $%
V_{0}=18.95$ [Figs. \ref{stability_N}(a$_{1}$-a$_{6}$) and (b$_{1}$-b$_{6}$%
), respectively] and the period of the radial potential $D=6.25$. In this
figure, stability areas are represented by orange bars in the plane of ($N,$O%
$_{\text{n}}$). It is seen that,\ for fixed values of $\left\vert
S_{1,2}\right\vert $ and $V_{0}$, the HV QDs are stable in finite intervals,
i.e., $N_{\min }<N<N_{\max }$. At $N<N_{\min }$, such modes do not exist,
while at $N>N_{\max }$ they spill out into the next radial trough or
multiple troughs.

\begin{figure}[tbp]
{\includegraphics[width=0.7\columnwidth]{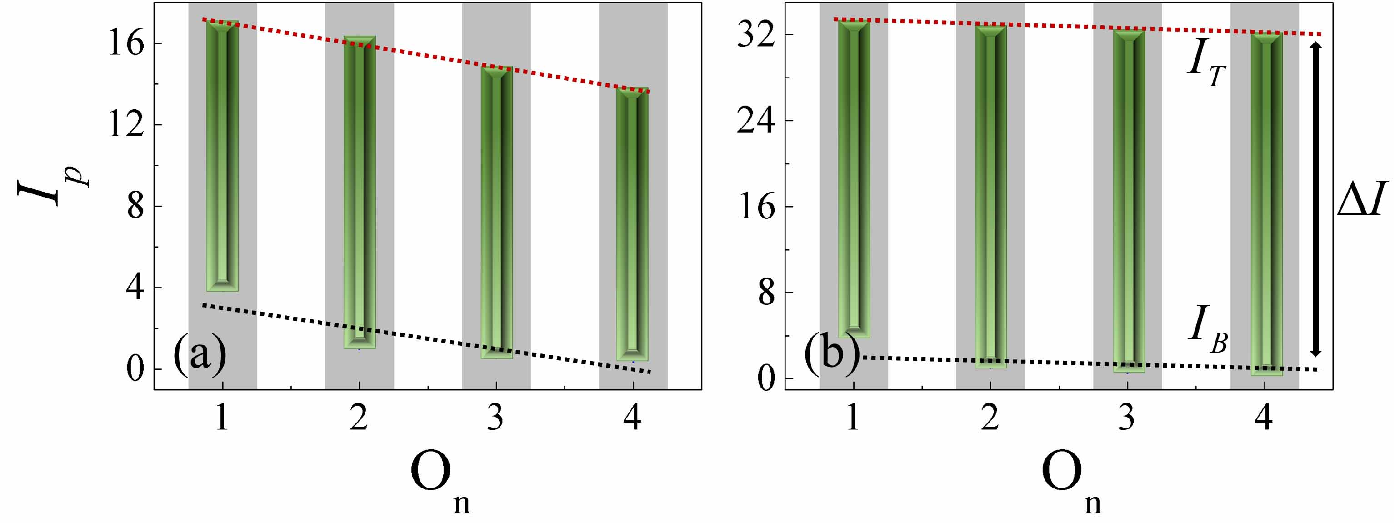}}
\caption{The peak density of the stable HV ring-shaped QDs with $S_{1,2}=\pm
1$, trapped in the radial troughs with O$_{\text{n}}=1,2,3$ and $4$. The
parameters are $D=6.25$ in both panels, and $V_{0}=12.63$ in (a), or $%
V_{0}=18.95$ in (b). The numerical results yield $\Delta
I_{1}=I_{T}-I_{B}\approx 14.09$ in (a) and $\Delta I_{2}=I_{T}-I_{B}\approx
31.29$ in (b).}
\label{density}
\end{figure}

The fact that the length of the stability interval in Fig. \ref{stability_N}
increases with the increase of O$_{\text{n}}$ can be explained as follows. The densities in the different O$_{\text{n}}$ are comparable,
then the total norm will be larger for higher $n$ due to the increase in the
size of the potential from the radial geometry. The peak-density difference
of the HV QDs in the stability region is%
\begin{equation}
\triangle I=I_{T}-I_{B},  \label{deltaI}
\end{equation}%
where $I_{T}$ and $I_{B}$ (represented by the red and black dotted lines,
respectively, in Fig. \ref{density}) are values of the peak density at the
top and bottom of the orange bars. The green bars in Fig. \ref{density}
illustrate the peak density of the stable HV ring-shaped QDs with $%
S_{1,2}=\pm 1$ trapped at O$_{\text{n}}=1,2,3$ and $4$, with $V_{0}=12.63$
in (a) and $V_{0}=18.95$ in (b), respectively. The numerical results yield $%
\Delta I_{1}\approx 14.09$ and $\Delta I_{2}\approx 31.29$ in Figs. \ref%
{density}(a) and (b), respectively. Different from ring-shaped QDs with
explicit, rather than hidden, vorticity trapped in radially-periodic
potentials \cite{NJP_LB2022}, the peak-density difference, $\Delta I$, of
the present modes substantially increases with the increase in $V_{0}$.

Further, comparing the above results for $V_{0}=12.63$ [Figs. \ref%
{stability_N}(a$_{1}$-a$_{6}$)] and $V_{0}=18.95$ [Figs. \ref{stability_N}(b$%
_{1}$-b$_{6}$)], we conclude that the stability intervals are shorter in the
former case. This finding can be explained too. The increase in the depth of
the potential, $V_{0}$, causes a slight decrease of the radial width $w$ of
the HV QDs, as can be seen from the cross section of $|\psi (x,0)|^{2}$ in
Fig. \ref{example}(a$_{3}$), where the red and blue dashed curves correspond
to $V_{0}=12.63$ and $V_{0}=18.95$, respectively. However, as said above,
the peak density difference, $\Delta I$, is larger in the latter case.
According to Ref. \cite{NJP_LB2022}, this results in increase of $\Delta N$.
It is also worthy to note that, as is shown by the comparison to Ref. \cite%
{NJP_LB2022}, the size of the stability intervals of the HV states is
smaller than their counterparts with the explicit vorticity, for equal
values of the norm and WNs.

The dependence of the chemical potential, $\mu $, of the HV QDs on the total
norm, $N$, is plotted in Fig. \ref{MU+D}(a), where parameters are the same
as in Fig. \ref{stability_N}(a$_{2}$). The $\mu (N)$ curves feature a
positive slope, $d\mu /dN>0$, violating the well-known Vakhitov-Kolokolov
(VK) criterion, which is a necessary stability condition for self-trapped
modes maintained by a self-attractive nonlinearity \cite{VK,Berge1998,book}.
In Fig. \ref{MU+D}(a), the solid and dotted lines represent values of $\mu $
of stable and unstable HV ring-shaped QDs, respectively. On the other hand,
for localized modes supported by self-repulsive nonlinearity (such as gap
soliton) the necessary stability condition is the opposite, given by the
anti-VK criterion, $d\mu /dN>0$ \cite{HS}.

In the case of the competition between the self-attraction and repulsion in
Eq. (\ref{GPE_2D}), it is not \textit{a priori} obvious which condition, VK
or anti-VK, is a dominant one for the stability. To resolve the issue, we,
first, employ a straightforward estimate for the peak density of the
ring-shaped modes,
\begin{equation}
I_{p}\approx N/[\pi (2O_{\text{n}}-1)Dw],  \label{IP}
\end{equation}%
where $w$ is the effective width of the HV QDs in the radial direction.
Next, following Ref. \cite{NJP_LB2022}, we apply the Thomas-Fermi (TF)
approximation to Eq. (\ref{radial_equation}), dropping the second-derivative
terms in it, which yields%
\begin{equation}
\mu _{\mathrm{TF}}=-\delta gI_{p}+2\sqrt{2}\Gamma I_{p}^{3/2}+\frac{1}{2}%
V_{0},  \label{mu_IP}
\end{equation}%
By combining this expression with Eq. (\ref{IP}), we conclude that $d\mu
/dN>0$ takes place above a threshold value,%
\begin{equation}
N>N_{\mathrm{th}}=\frac{\pi (2\text{O}_{\text{n}}-1)Dw\left( \delta g\right)
}{18\Gamma ^{2}},  \label{nth}
\end{equation}%
In our computations, the radial period is fixed as $D=6.25$, and if we
select $w=D/2$, then the $N_{\mathrm{th}}$ values in Eq. (\ref{nth}) for O$_{%
\text{n}}=1,2,3,4$ and $5$ are $N_{\mathrm{th}}\approx
4.59,13.77,22.94,32.12 $ and $41.3$, respectively. On the other hand, it is
seen in Fig. \ref{MU+D}(a) that numerically found bottom stability
boundaries for O$_{\text{n}}=1,2,3,4$ and $5$ are $N_{\min }\approx
508,343,304,282$ and $262$, respectively, which are all much larger than $N_{%
\mathrm{th}}$ for each value of O$_{\text{n}}$, hence the anti-VK criterion
indeed determines the stability in this case.

The multistability of the HV ring-shaped QDs, which may coexist as stable
modes with the same norm, trapped in the troughs with different numbers O$_{%
\text{n}}$, is illustrated by the vertical dashed line in Fig. \ref{MU+D}%
(a). In this case, it is relevant to compare energies of the coexisting
modes. The Hamiltonian density corresponding to Eq. (\ref{GPE_2D}) is
\begin{eqnarray}
\mathcal{H} &=&\frac{1}{2}\left( \left\vert \nabla \psi _{1}\right\vert
^{2}+\left\vert \nabla \psi _{2}\right\vert ^{2}\right) +\frac{g}{2}\left(
\left\vert \psi _{1}\right\vert ^{4}+\left\vert \psi _{2}\right\vert
^{4}\right) -\left( g+\delta g\right) \left\vert \psi _{1}\right\vert
^{2}\left\vert \psi _{2}\right\vert ^{2}  \notag \\
&&+\frac{2\Gamma }{5}\left( \left\vert \psi _{1}\right\vert ^{2}+\left\vert
\psi _{2}\right\vert ^{2}\right) ^{5/2}+V_{0}\cos ^{2}\left( \frac{\pi }{D}%
r\right) \left( \left\vert \psi _{1}\right\vert ^{2}+\left\vert \psi
_{2}\right\vert ^{2}\right) .  \label{La}
\end{eqnarray}%
Using Eq. (\ref{La}), one can find that, among the HV QDs with equal norms
and equal values of $\left\vert S_{1,2}\right\vert $, the minimum energy, $%
\int \int \mathcal{H}dxdy$, corresponds to largest ring's number O$_{\text{n}%
}$.

The effect of the period of the radial potential, $D$, on the HV QDs is
presented in Fig. \ref{MU+D}(b), where the other parameters are fixed as $%
N=1500$, $S_{1,2}=\pm 2$, and $V_{0}=12.63$. Our results demonstrate that
the modes are stable in certain intervals, $D_{\min }<D<D_{\max }$, for
given ring numbers O$_{\text{n}}$, \textit{viz}.,
\begin{equation}
\Delta D\equiv D_{\max }-D_{\min }\varpropto \frac{1}{(2\text{O}_{\text{n}%
}-1)}.  \label{DeltaD}
\end{equation}%
This finding may be explained, in terms of the above-mentioned modulational
instability, by attenuation of the stabilizing effect of the ring's
curvature, which scales as $(2$O$_{\text{n}}-1)^{-1}$.

\begin{figure}[tbp]
{\includegraphics[width=0.75\columnwidth]{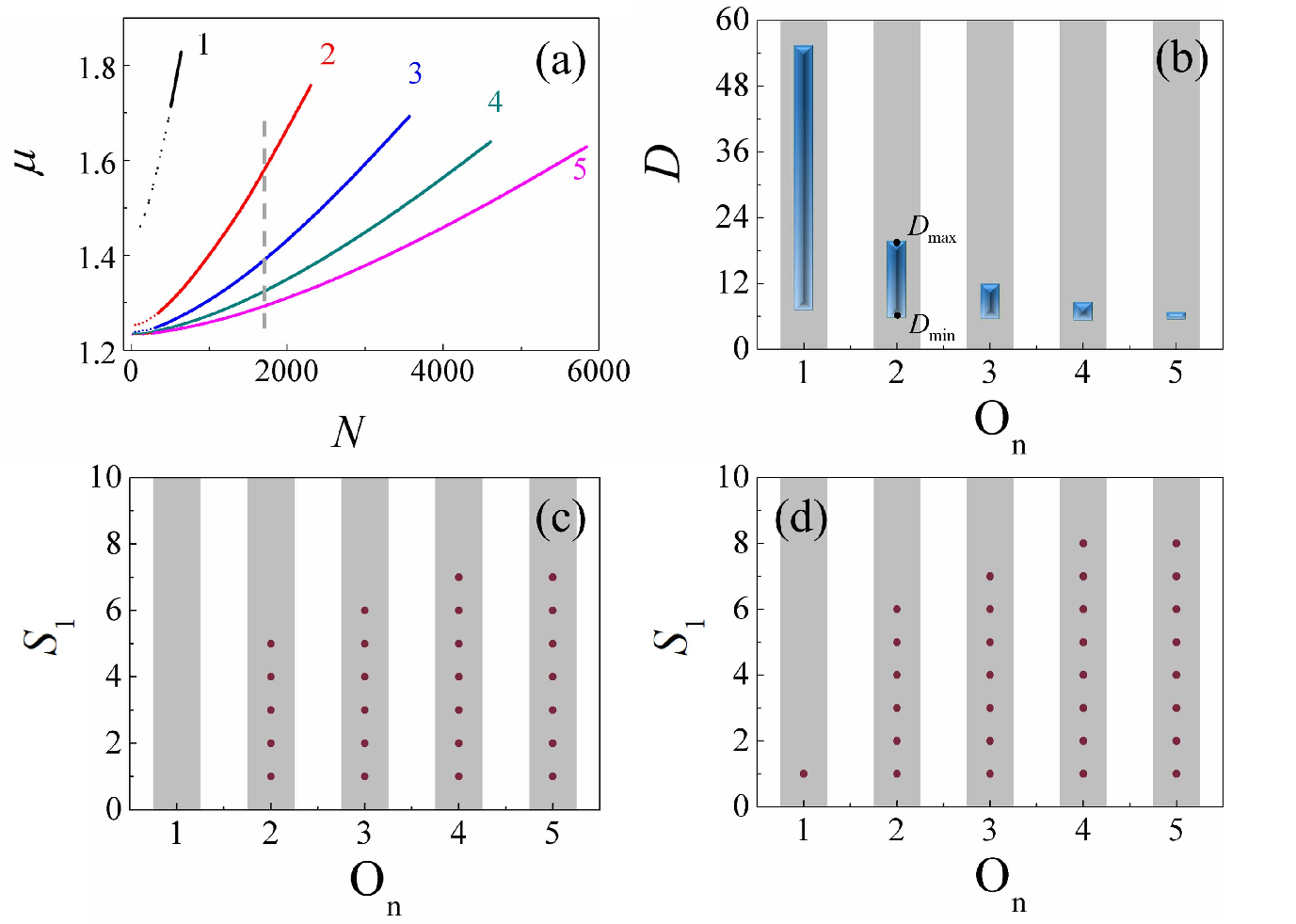}}
\caption{(a) The dependence of the chemical potential, $\protect\mu $, on
norm, $N$, for HV ring-shaped QDs with $S_{1,2}=\pm 2$, where the parameters
are the same as in Fig. \protect\ref{stability_N}(a$_{2}$). Solid and dotted
segments\ represent stable and unstable QDs, respectively. The vertical
dashed line illustrates the coexitence of stable modes with equal norms
which are trapped in the troughs with different numbers O$_{\text{n}}$. (b)
Boundary values ($D_{\min }$ and $D_{\max }$) of the period of the radial
potential for given values of O$_{\text{n}}$, between which the HV
ring-shaped QDs are stable, the other parameters being $N=1500$, $%
V_{0}=12.63 $, and $S_{1,2}=\pm 2$. Panels (c) and (d) show the trapping
ability of the radial-potential troughs for HV ring-shaped QDs with
indicated values of WNs $\left\vert S_{1,2}\right\vert $ in the radial
troughs with numbers O$_{\text{n}}$ and $V_{0}=12.63$ (c) or $V_{0}=18.95$
(d), the other parameters being $N=1500$ and $D=6.25$. }
\label{MU+D}
\end{figure}

With the norm fixed to $N=1500$ and the radial period fixed to $D=6.25$, the
ability of the radial potential (\ref{V_2D}) to maintain the stable HV
ring-shaped QDs with high values of WN, $\left\vert S_{1,2}\right\vert $, at
different numbers of O$_{\text{n}}$ is illustrated in Figs. \ref{MU+D}(c)
and (d) for $V_{0}=12.63$ and $V_{0}=18.95$, respectively. The results are
consistent with those presented in Fig. \ref{stability_N}(a$_{1}$) and (b$%
_{1}$). For $V_{0}=12.63$, no stable solutions are found at O$_{\text{n}}=1$%
, which is explained by the fact that the above-mentioned value, $N_{\max
}=838$, is smaller than the fixed norm adopted here, $N=1500$, for $%
S_{1,2}=\pm 1$ at O$_{\text{n}}=1$. On the other hand, the HV ring-shaped
QDs with $S_{1,2}=\pm 1$ at O$_{\text{n}}=1$ are found for $V_{0}=18.95$,
because in that case we have $N_{\min }=153<1500<N_{\max }=1695$. In Figs. %
\ref{MU+D}(c) and (d), one can see that the holding capacity of the radial
troughs increases with the increase of O$_{\text{n}}$, and these plots also
reveal that the increase of the depth of the potential naturally enhances
the trough's capacity to hold stable HV\ QDs.

\begin{figure}[tbp]
{\includegraphics[width=0.8\columnwidth]{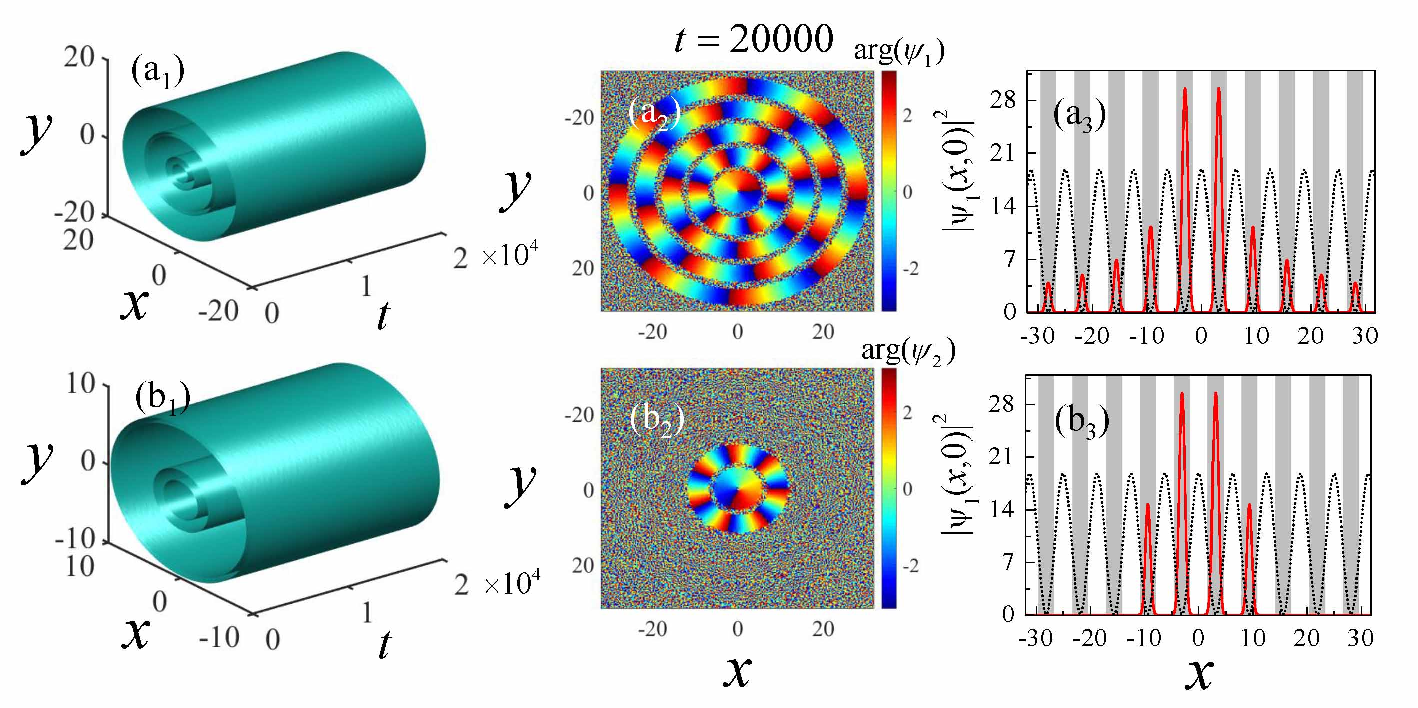}}
\caption{Typical examples of stable multiring (nested) HV QDs. (a$_{1}$)
Direct simulations of the evolution of the $\protect\psi _{1}$ component for
a stable multiring state with parameters $(N,S_{1,2}$,O$_{\text{n}})$ $%
=(1500,\pm 1,1)$, $(1500,\pm 6,2)$, $(1500,\pm 7,3)$, $(1500,\pm 8,4)$, and $%
(1500,\pm 8,5)$ (a$_{2}$) The output phase pattern of the same state at $%
t=20000$. (a$_{3}$) The cross section along $y=0$, $|\protect\psi %
_{1,2}(x,0)|^{2}$, plotted by the red curves. (b$_{1}$) Direct simulations
of the evolution of the $\protect\psi _{1}$ component for a stable two-ring
HV mode with opposite signs of $S_{1,2}$ in the rings, with parameters $%
(N,S_{1,2}$,O$_{\text{n}})=$ $(1500,\pm 1,1)$ and $(2000,\mp 7,2)$. (b$_{2}$%
) The output phase pattern of the same mode at $t=20000$. (b$_{3}$) The
cross section along $y=0$, $|\protect\psi _{1,2}(x,0)|^{2}$, plotted by red
curves. Other parameters are the same as in Fig. \protect\ref{MU+D}(d).}
\label{nest_QDs}
\end{figure}

\section{Nested HV quantum droplets}

Nested structures are stable concentric states in which self-trapped vortex
rings carrying different WNs (topological charges) are embedded into each
other. This concept was originally introduced for dissipative ring-vortex
solitons \cite{Skarka}. Later, nested multi-ring states with different WNs
trapped in different radial troughs were produced \cite{NJP_LB2022}. The
results of those studies demonstrate that mutually embedded ring-shaped
vortical may be stable if the separation between the rings is large enough.

Typical examples of stable concentric HV states are displayed in Fig. \ref%
{nest_QDs}, where panel (a$_{1}$) shows the result of direct simulations for
the $\psi _{1}$ component, which is composed of five concentric HV QDs with (%
$N$,$S_{1,2}$,O$_{\text{n}})$ $=$ $(1500,\pm 1,1)$, $(1500,\pm 6,2)$, $%
(1500,\pm 7,3)$, $(1500,\pm 8,4)$, and $(1500,\pm 8,5)$, at $t=0$ and $20000$%
. This mode stays stable during time exceeding $100$ ms in physical units.
Its phase pattern at $t=20000$ is shown in Fig. \ref{nest_QDs}(a$_{2}$), and
the density cross section, $|\psi _{1}(x,0)|^{2}$, is displayed in Fig. \ref%
{nest_QDs}(a$_{3}$).

It is worthy to consider the case of nested states composed of ring-shaped
HV QDs with \emph{opposite signs }of WNs. A typical example of a stable
concentric mode of this type, with parameters of ($N$,$S_{1,2}$,O$_{\text{n}%
})$ $=$ $(1500,\pm 1,1)$ and $(2000,\mp 7,2)$, are displayed in Figs. \ref%
{nest_QDs}(b$_{1}$). Additionally, the output phase pattern and density
cross section, $|\psi _{1}(x,0)|^{2}$, are displayed in Figs. \ref{nest_QDs}%
(b$_{2}$) and (b$_{3}$).

In this work, the radially-periodic potential can maintain self-trapping of
HV ring-shaped QDs in particular troughs. If the QDs are stable in their
troughs, the corresponding nested patterns are also stable, the reason being
that the HV rings trapped in different troughs are nearly isolated from each
other. These results offer a potential use in the design of data-storage
devices, in which different data components, coded by the respective WN
values, may be deposited in different radial troughs.

\section{Conclusion}

The purpose of this work is to establish stability and characteristics of 2D
ring-shaped QDs (quantum droplets) with HV (hidden vorticity) formed by
binary BEC in the radially-periodic potential. The system is modeled by the
coupled GP equations with the LHY (Lee-Huang-Yang) terms, which represent
the correction to the MF (mean-field) theory produced by quantum
fluctuations, and the radial potential. Families of ring-shaped QDs with the
HV structure, represented by high values of the WN (winding number), trapped
in particular circular troughs of the radial potential, have been produced.
Effects of the depth and period of the radial potential on the ring-shaped
HV QDs were studied, and their stability area was identified. It was found
that the size of the stability interval increases with the increase of the
trough's number, O$_{\text{n}}$. On the other hand, the stability interval
naturally expands with the increase of the potential's depth. For stable HV
ring-shaped QDs, the dependence between the chemical potential and total
norm obeys the anti-VK criterion, which is explained by the dominant role of
the self-repulsive LHY nonlinearity. The multistability of the HV states was
demonstrated, in the form of the coexistence of the modes with equal norms
and WNs but different radii of the trapping circular trough. The trapping
capacity of the troughs was identified for the HV ring-shaped QDs with
different WNs. Nested patterns composed of rings with different WNs, trapped
in different radial troughs, were constructed, including two-ring patterns
with opposite WN signs. The results reported in this work suggest a new
approach to the creation of stable nested QDs with embedded WNs. In
particular, these results may also be used in the design of data-storage
devices, with stable WN rings trapped in different radial troughs encoding
data components.

The analysis can be extended in other directions -- in particular, one can
consider ring-shaped QDs of the HV type in an elliptically deformed
potential. A challenging possibility is to the extend the consideration for
three-dimensional settings.

\section{Acknowledgments}

This work was supported by Natural Science Foundation of Guangdong province
through grant No. 2021A1515010214, NNSFC (China) through grants Nos.
12274077, 11905032, 12274077, and 11904051, the Guangdong Basic and Applied
Basic Research Foundation through grants Nos. 2019A1515110924 and
2021A1515111015, the Key Research Projects of General Colleges in Guangdong
Province through grant No. 2019KZDXM001, the Special Funds for the
Cultivation of Guangdong College Students Scientific and Technological
Innovation through grants Nos. pdjh2021b0529 and pdjh2022a0538, the Research
Fund of the Guangdong-Hong Kong-Macao Joint Laboratory for Intelligent
Micro-Nano Optoelectronic Technology through grant No. 2020B1212030010, and
Israel Science Foundation through Grant No. 1695/22.

\section{Appendix}

The matrix elements, $A_{11}\sim A_{44}$, in the linearized BdG equations (%
\ref{eig_eq}) are

\begin{align}
A_{11}& ={-}\frac{1}{2}\left( {\frac{d}{rdr}}+\frac{d^{2}}{dr^{2}}-{\frac{%
\left( S+m\right) ^{2}}{r^{2}}}\right)  \notag \\
& +\left( 2g|\phi _{1}|^{2}-\left( g+\delta g\right) |\phi _{2}|^{2}+\frac{3%
}{2}\Gamma \left( |\phi _{1}|^{2}+|\phi _{2}|^{2}\right) ^{1/2}|\phi
_{1}|^{2}+\Gamma \left( |\phi _{1}|^{2}+|\phi _{2}|^{2}\right) ^{3/2}+V-\mu
\right) ,  \notag \\
A_{12}& =g\phi _{1}^{2}+\frac{3}{2}\Gamma \left( |\phi _{1}|^{2}+|\phi
_{2}|^{2}\right) ^{1/2}\phi _{1}^{2},  \notag \\
A_{13}& =\frac{3}{2}\Gamma \left( |\phi _{1}|^{2}+|\phi _{2}|^{2}\right)
^{1/2}\phi _{1}\phi _{2}^{\ast }-\left( g+\delta g\right) \phi _{1}\phi
_{2}^{\ast },  \notag \\
A_{14}& =\frac{3}{2}\Gamma \left( |\phi _{1}|^{2}+|\phi _{2}|^{2}\right)
^{1/2}\phi _{1}\phi _{2}-\left( g+\delta g\right) \phi _{1}\phi _{2},
\label{A1}
\end{align}%
\begin{align}
A_{21}& =-g\phi _{1}^{\ast 2}-\frac{3}{2}\Gamma \left( |\phi _{1}|^{2}+|\phi
_{2}|^{2}\right) ^{1/2}\phi _{1}^{\ast 2},  \notag \\
A_{22}& =\frac{1}{2}\left( {\frac{d}{rdr}}+\frac{d^{2}}{dr^{2}}-{\frac{%
\left( S-m\right) ^{2}}{r^{2}}}\right)  \notag \\
& +\left( g+\delta g\right) |\phi _{2}|^{2}-2g|\phi _{1}|^{2}-\frac{3}{2}%
\Gamma \left( |\phi _{1}|^{2}+|\phi _{2}|^{2}\right) ^{1/2}|\phi
_{1}|^{2}-\Gamma \left( |\phi _{1}|^{2}+|\phi _{2}|^{2}\right) ^{3/2}-V+\mu ,
\notag \\
A_{23}& =\left( g+\delta g\right) \phi _{1}^{\ast }\phi _{2}^{\ast }-\frac{3%
}{2}\Gamma \left( |\phi _{1}|^{2}+|\phi _{2}|^{2}\right) ^{1/2}\phi
_{1}^{\ast }\phi _{2}^{\ast },  \notag \\
A_{24}& =\left( g+\delta g\right) \phi _{1}^{\ast }\phi _{2}-\frac{3}{2}%
\Gamma \left( |\phi _{1}|^{2}+|\phi _{2}|^{2}\right) ^{1/2}\phi _{1}^{\ast
}\phi _{2},  \label{A2}
\end{align}%
\begin{align}
A_{31}& =\frac{3}{2}\Gamma \left( |\phi _{2}|^{2}+|\phi _{1}|^{2}\right)
^{1/2}\phi _{2}\phi _{1}^{\ast }-\left( g+\delta g\right) \phi _{2}\phi
_{1}^{\ast },  \notag \\
A_{32}& =\frac{3}{2}\Gamma \left( |\phi _{2}|^{2}+|\phi _{1}|^{2}\right)
^{1/2}\phi _{2}\phi _{1}-\left( g+\delta g\right) \phi _{2}\phi _{1},  \notag
\\
A_{33}& ={-}\frac{1}{2}\left( {\frac{d}{rdr}}+\frac{d^{2}}{dr^{2}}-{\frac{%
\left( -S+m\right) ^{2}}{r^{2}}}\right)  \notag \\
& +2g|\phi _{2}|^{2}-\left( g+\delta g\right) |\phi _{1}|^{2}+\frac{3}{2}%
\Gamma \left( |\phi _{2}|^{2}+|\phi _{1}|^{2}\right) ^{1/2}|\phi
_{2}|^{2}+\Gamma \left( |\phi _{2}|^{2}+|\phi _{1}|^{2}\right) ^{3/2}+V-\mu ,
\notag \\
A_{34}& =g\phi _{2}^{2}+\frac{3}{2}\Gamma \left( |\phi _{2}|^{2}+|\phi
_{1}|^{2}\right) ^{1/2}\phi _{2}^{2},  \label{A3}
\end{align}%
\begin{align}
A_{41}& =\left( g+\delta g\right) \phi _{1}^{\ast }\phi _{2}^{\ast }-\frac{3%
}{2}\Gamma \left( |\phi _{1}|^{2}+|\phi _{2}|^{2}\right) ^{1/2}\phi
_{1}^{\ast }\phi _{2}^{\ast },  \notag \\
A_{42}& =\left( g+\delta g\right) \phi _{2}^{\ast }\phi _{1}-\frac{3}{2}%
\Gamma \left( |\phi _{1}|^{2}+|\phi _{2}|^{2}\right) ^{1/2}\phi _{2}^{\ast
}\phi _{1},  \notag \\
A_{43}& =-g\phi _{2}^{\ast 2}-\frac{3}{2}\Gamma \left( |\phi _{2}|^{2}+|\phi
_{1}|^{2}\right) ^{1/2}\phi _{2}^{\ast 2},  \notag \\
A_{44}& =\frac{1}{2}\left( {\frac{d}{rdr}}+\frac{d^{2}}{dr^{2}}-{\frac{%
\left( -S-m\right) ^{2}}{r^{2}}}\right)  \notag \\
& +\left( g+\delta g\right) |\phi _{1}|^{2}-2g|\phi _{2}|^{2}-\frac{3}{2}%
\Gamma \left( |\phi _{2}|^{2}+|\phi _{1}|^{2}\right) ^{1/2}|\phi
_{2}|^{2}-\Gamma \left( |\phi _{2}|^{2}+|\phi _{1}|^{2}\right) ^{3/2}-V+\mu
_{{}}.  \label{A4}
\end{align}

\end{document}